# THE IMPACT OF PROPAGATION ENVIRONMENT AND TRAFFIC LOAD ON THE PERFORMANCE OF ROUTING PROTOCOLS IN AD HOC NETWORKS

A. Rhattoy[1] and A. Zatni[2]

[1]Modeling, Systems and Telecommunications Research Group, Department of Computer, Moulay Ismail University, Higher School of Technology, B. P. 3103, 50000,Toulal, Meknes, Morocco
[2]Optronics Laboratory, Department of Computer, Ibnou Zohr University, Higher School of Technology, B. P. 33/S, 80000, Agadir, Morocco
rhattoy@gmail.com and zatni@esta.ac.ma

## ABSTRACT

Wireless networks are characterized by a dynamic topology triggered by the nodes mobility. Thus, the wireless multi-hops connection and the channel do not have a determinist behaviour such as: interference or multiple paths. Moreover, the nodes' invisibility makes the wireless channel difficult to detect. This wireless networks' behaviour should be scrutinized. In our study, we mainly focus on radio propagation models by observing the evolution of the routing layer's performances in terms of the characteristics of the physical layer. For this purpose, we first examine and then display the simulation findings of the impact of different radio propagation models on the performance of ad hoc networks. To fully understand how these various models influence the networks performance, we have compared the performances of several routing protocols (DSR, AODV, and DSDV) for each propagation model. To complete our study, a comparison of energy performance based routing protocols and propagation models are presented. In order to reach credible results, we focused on the notion of nodes' speed and the number of connections by using the well known network simulator NS-2.

## KEYWORDS

*Mobile Ad-hoc, Routing Protocols, Fading, Propagation Model, NS-2.* Network Lifetime, Energy Consumption.

## 1. INTRODUCTION

Before using a wireless network or installing the stations of a cellular network, we have to determine the radio waves' targeted coverage. The targeted radio coverage has a crucial economic impact because it determines the equipment to be utilized. In other words, the bigger the coverage is, the less antennas are required to cover the region or to reach a grand area. Besides, the radio coverage depends on several parameters such as the emission power. However, the environment where the waves spread and the utilized frequency also play a crucial role. The radio propagation waves are controlled by strict rules, mainly when there are obstacles between the transmitter and the receiver [1], [2]. Among the changes a wave may undergo, we can cite: reflection, diffraction, diffusion and absorption (figure 1). The metrics used are packet delivery fraction, delay, throughput and energy. The remainder of the paper is outlined as follows: Section (2) focuses on the radio propagation models types. Section (3) discusses of routing protocols concepts in ad hoc networks. In Section (4) the methodologies of simulation are introduced. Section (5), we investigate the impact of radio propagation models on the performances of routing protocols in ad hoc networks and the energy consumption. Finally, we present our conclusions in Section (6).

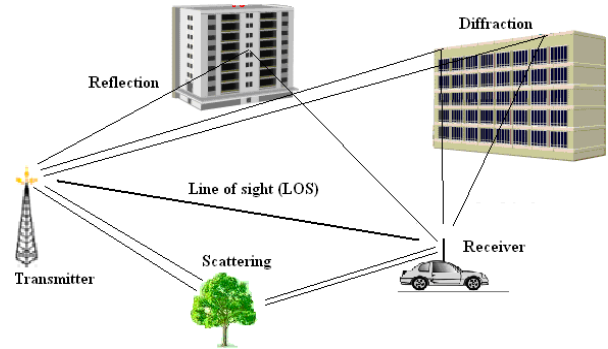

Figure 1.  The different physical phenomena disturbing radio signal propagation

## 2. RADIO PROPAGATION MODELS

In a propagation model, we use a set of mathematical models which are supposed to provide an increasing precision. Propagation radio models are three types: path loss, shadowing and fading. The first type can be expressed as the power loss during the signal propagation in the free space. The second type is characterized by fixed obstacles on the path of the radio signal propagation. The third category is the fading which is composed of multiple propagation distances, the fast movements of transmitters and receivers units and finally the reflectors [3].

### 2.1. Free Space Model

The free space model assumes that in the ideal propagation condition between the transmitter and the receiver, there is only one clear line of slight (LOS) path. The following equation calculates the received signal power in a free space with distance (d) from the sender:

$$P_r(d) = \frac{P_t G_t G_r \lambda^2}{(4\pi)^2 d^2 L} \qquad (1)$$

Where $P_t$ is the power transmission (in watts), $G_t$ and $G_r$ are the antenna gains of the transmitter and receiver respectively. L is the system loss factor. λ is the wave length and d is the distance between the transmitter and the receiver [4], [5].

### 2.2. Two-Ray Ground Model

The free space model mentioned above states that there is only one single direct path. In fact, the signal reaches the receiver through multiple paths (due to reflection, refraction and scattering). The two-path model attempts to account for this phenomenon. In other words, the model advocates that the signal attains the receiver via true paths: a line-of-slight path and a path through which the reflected wave is received [6]. In the two-path model, the received power is represented by:

$$P_r(d) = \frac{P_t G_t G_r h_t^2 h_r^2}{d^4 L} \qquad (2)$$

Where $h_t$ and $h_r$ are the heights of the transmitter and receiver respectively. Nonetheless, for short distances, the two-ray model does not give accurate results because of in oscillation caused by the constructive and destructive combination of the two rays. The propagation model in the free space is instead, still used where d is small. Hence, in this model, we calculate $d_c$ as a cross-over distance. When $d < d_c$, we use the first equation, but when $d > d_c$, the second equation is used. At the cross-over distance, equations (1) and (2) give similar results. Consequently, $d_c$ can be calculated as:

$$d_c = \frac{4\pi d_t h_r}{\lambda} \qquad (3)$$

## 2.3 Shadowing Model

Both the free space and the two-ray models predict the received power in terms of the distance. They also represent a communication area as an ideal circle. In fact, the received power at a given distance varies randomly because of multi-path propagation effects, known as fading effects. Thus, the two aforementioned models predict the mean received power at distance d. The shadowing model is two fold [7]. The first model is the path loss model represented by $P_r(d)$. It employs a close in distance $d_0$ as follows:

$$\frac{P_r(d_0)}{P_r(d)} = \left(\frac{d}{d_0}\right)^{\beta} \qquad (4)$$

$\beta$ is called the path loss exponent, and is often empirically determined by filed measurement. Equation (4) implies that $\beta = 2$ in free space propagation. The table.1 gives typical values of $\beta$ [8].

Table 1: Some Typical values of path loss $\beta$

| Environment | | $\beta$ |
|---|---|---|
| Outdoor | Free space | 2 |
| | Shadowed urban area | 2.7 to 5 |
| In building | Line-of-sight | 1.6 to 1.8 |
| | Obstructed | 4 to 6 |

Langer values of $\beta$ correspond to more obstructions and thus faster decrease in average received power as distance becomes larger. From equation (4), we have:

$$\left[\frac{P_r(d)}{P_r(d_0)}\right]_{dB} = -10\beta \log\left(\frac{d}{d_0}\right) \qquad (5)$$

The second part of the shadowing model reflects the variations of received power at certain distance. It is a log-normal random variable. The overall model is represented by:

$$\left[\frac{P_r(d)}{P_r(d_0)}\right]_{dB} = -10\beta \log\left(\frac{d}{d_0}\right) + X_{dB} \qquad (6)$$

Where $X_{dB}$ is Gaussian random variable with zero mean and standard deviation $\sigma$dB. $\sigma$dB is called shadowing deviation, and also obtained through measurement in the real environment. Table 2 displays some typical values of $\sigma$dB. This equation is also labelled a log-normal shadowing model.

Table 2: Typical values of shadowing deviation $\sigma$dB

| Environment | $\sigma_{dB}$ (dB) |
|---|---|
| Outdoor | 4 to 12 |
| Office, hard partition | 7 |
| Office, soft partition | 9.6 |
| Factory, line-of-sight | 3 to 6 |
| Factory, obstructed | 6.8 |

## 2.4. Small-Scale Fading model: Rayleigh and Rice

This fading model depicts the rapid fluctuations of the received signal due to multipath fading. This fading phenomenon is generated by the interference of at least two types of transmitted signals to the receiver with slight time intervals [9], [10]. The outcome may vary according to fluctuations and to different phases in terms of multiple factors such as: delay between waves, the intensity and the signal band width. Hence, the system performance may be attenuated by the fading. However, there are several techniques that help stopping this fading. The signal fading were monitored according to a statistical law wherein the most frequently used distribution is Raleigh's [11]. The transmitted signal is, thus, conditioned by the following phenomena: reflection, scattering and diffusion. Thanks to these three phenomena, the transmitted power may reach the hidden areas despite the lack of direct visibility (NLOS) between the transmitter and receiver. Consequently, the amount of the received signal has a density of Rayleigh:

$$f(x) = \begin{cases} \dfrac{2x}{P} \exp(-\dfrac{x^2}{P}) \text{,} & pour\ 0 \leq x \leq \infty \\ 0 & , pour\ x < 0 \end{cases} \tag{7}$$

Where, P is the average received power. In case where there is a direct path (LOS) between the transmitter and receiver, the signal no longer obeys to Rayleigh's law but to Rice's. The probability density of Rice is represented by:

$$f(x) = \begin{cases} \dfrac{2x(K+1)}{P} \exp\left(-K - \dfrac{(K+1)x^2}{P}\right) I_0 \left(2x\sqrt{\dfrac{K(K+1)}{P}}\right) \text{,} & pour\ 0 \leq x \leq \infty \\ 0 & , pour\ x < 0 \end{cases} \tag{8}$$

Where K is the ratio of the power received in the direct line and in the path, P is the average power received and $I_0(x)$ is the zero-order Bessel function de fined by:

$$I_0(x) = \frac{1}{2\pi} \int_0^{2\pi} \exp(-x\cos\theta)d\theta \text{,} \tag{9}$$

The density of Rice (Equation 8) is reduced to the density of Rayleigh (Equation 7) in the case of an absence of a direct path which means that K=0 and thus $I_0(x) = 1$.

### 2.5. Nakagami model

This distribution encompasses several other distributions as particular cases. To describe Rayleigh distribution, we assumed that the transmitted signals are similar and their phases are approximate. Nakagami model is more realistic in that it allows similarly to the signals to be approximate. Since we have used the same labels as in Rayleigh and Rice cases, we have $r = \left| \sum r_i e^{j\theta_i} \right|$. The probability density of Nakagami related to r is represented by:

$$P_r(r) = \frac{2m^m r^{2m-1}}{\Gamma(m)\Omega^m} \exp\left(-\frac{mr^2}{\Omega}\right) \text{,} \quad r \geq 0 \tag{10}$$

Where $\Gamma(m)$ is gamma function, $\Omega = E(r^2)$ and $m = \{E(r^2)\}^2 / Var(r^2)$ with the constraint $m \geq \frac{1}{2}$. Nakagami model is a general distribution of fading which is reduced to Rayleigh's distribution for m=1 and to unilateral Gaussian model for m=1/2. Besides, it represents pretty much rice model and it is closer to certain conditions in the lognormal distribution [12], [13] and [14].

## 3. AD HOC ROUTING PROTOCOLS

Ad hoc routing protocols are based on fundamental principles of routing such as: Inundation (flooding), the distance Vector, the routing to the source and the state of the site. According to the way routes are created and maintained during the data delivery, the routing protocols can be characterised into two categories: proactive and reactive [15]. Among the tested protocols in this study, only DSDV is proactive and the others (DSR and AODV) are all reactive. Proactive protocols update route information periodically, whereas reactive protocols establish routes only when needed. Here is a summary of the routing protocols assessed in this paper.

### 3.1. Dynamic Source Routing (DSR)

During the discovery process of routing, a source node generates a route-request packet which needs a new route to a certain destination. The route request is connected through the network until it reaches some nodes with a route to destination. A reply packet containing all information of intermediate nodes is sent back to the source. The sent packets contain a list of all nodes through which they have to transit. This list can be huge in a network with a big diameter. The nodes do not need the routing table. There are two DSR basic operations: the route discovery and the route maintenance. In order to cut down the expenses and the frequency of the route discovery, every single node keeps track of the paths thanks to reply packets. These paths are used until they become useless [16].

### 3.2. Ad-hoc On-Demand Distance Vector protocol (AODV)

AODV has a way for route request close to that of DSR. However, AODV does not perform a routing to the source. Every single node on the path refers to a point towards its neighbour from which it receives a reply. When a transit node needs broadcasts a route request to a neighbour, it also stores the node identifier in the routing table from which the first reply is received. To check the links state, AODV uses control messages (Hello) between direct neighbours. Besides, AODV utilizes a sequence number to avoid a round trip and to ensure using the most recent routes [16].

### 3.3. DSDV Protocol

The algorithm DSDV (Dynamic destination Sequenced Distance Vector) [16] has been constructed for mobile networks. Each mobile station keeps a routing table which contains all possible destinations, number of hops to reach the destination, sequence number (SN) associated with the node destination to distinguish the new routes of the old a ones and avoid the formation of round trip routing. The table updating is periodically transmitted across the network so as to sustain the information consistency, and thus generates an important traffic.

## 4. METHODOLOGY

In this paper, we have compared several routing protocols performances (DSR, AODV, and DSDV) according to every propagation models, focusing on their performance of energy consumption. In order to obtain valid results, we have inserted the notion of the number of connections. So as to analyse the ad hoc routing protocols' behaviour, we selected traffic sources with a constant output (CBR) related to UDP protocol. The packet emission rate is settled at 8 packets per second. We display the impact of the traffic load on the routing protocols. For this reason, we have varied a number of connections. Six cases were considered: 5, 10, 15, 20, 25 and 30 connections. The assessed protocols are: AODV, DSR and DSDV. These three are available in 2.34 of ns-2. The propagation models under study are: the free space, the two-Ray ground, shadowing, Rice's and Nakagami's models. The simulation span is of 200 seconds. The data packet size is 512 octets. The mobile nodes utilize the random waypoint mobility model [17]. The Mobil nodes move within a square dimension area 670mx 670m. For the time being, let's limit the nodes' maximal speed at 5m.s-1. Because of the length chosen in this paper, we have selected just three performance indicators in order to study the routing protocols performances. They are outlined as follows: Packet Delivery Fraction, end Average to end delay, the throughput and residual energy consumption.

***Packet Delivery Fraction (PDF):*** This is the ratio of total number of CBR packets successfully received by the destination nodes to the number of CBR packets sent by the source nodes throughout the simulation.

$$Pkt\_Delivery\% = \frac{\sum_{1}^{n} CBR_{recv}}{\sum_{1}^{n} CBR_{sent}} \times 100$$

This estimate gives us an idea of how successful the protocol is in delivering packets to the application layer. A high value of PDF indicates that most of the packets are being delivered to the higher layers and is a good indicator of the protocol performance.

***Average end to end delay (AE2E Delay):*** This is defined as the average delay in transmission of a packet between two nodes and is calculated as follows:

$$Avg\_End\_to\_End\_delay = \frac{\sum_{1}^{n} \left( CBR_{sent\_Time} - CBR_{recv\_Time} \right)}{\sum_{1}^{n} CBR_{recv}}$$

A higher value of end-to-end delay means that the network is congested and hence the routing protocol doesn't perform well. It depends on the physical characteristics of a link, and the delay of treatment and the state of the queues of the nodes.

***Throughput:***
The throughput data reflects the effective network capacity. It is computed by dividing the message size with the time it took to arrive at its destination. It is measured considering the hops performed by each packet

***Energy consumption model:***
Because of the crucial importance power management in mobile ad hoc networks, it is interesting to evaluate the energy consumption induced by each of the protocols studied. Recall that because of their mobility, the terminals used in MANETs get their energy (relatively limited) embedded batteries. The Energy Model, as implemented in NS-2, is a node attribute. The energy model in a node has an initial value corresponding to the node energy level at the beginning of the simulation. It also takes into account energy consumption associated to each packet transmission and reception. When the node energy level goes down to zero, the node dies out, no more packets can be received or transmitted by the node. The energy consumption model which use in this research is adopted from [18]. Energy is converted in joules by multiplying power with time. The following equations are used to convert energy in joules:
Transmitted Energy:

$$T_x \, Energy = (T_x \, power \, * \, Packet \, Size) / 2 * 10^6$$

Receiving Energy:

$$R_x \, Energy = (R_x \, power \, * \, Packet \, Size) / 2 * 10^6$$

Total energy consumed by each node during transmission and reception is calculated by the following equation:

$$Total \, Energy \, Consumed = Initial \, energy - Energy \, left \, at \, each \, node$$

# 5. SIMULATION FINDINGS

The results corresponding to the PDF, AE2E Delay, Throughput and Energy consommé are shown in Figure 2, Figure 3, Figure 4, and Figure 5, respectively.

## 5.1. Packet Delivery Fraction (PDF)

Figure 2 displays, different routing protocols performances in terms of the number of connections. The charts also display that if the number of connections increases, the delivery fraction value tends to decrease for all models. Thus, there is network congestion. In this scenario, DSDV is less preferment than AODV and DSR because their PDF are over 99% in so far as it reaches 10 connections. However, when we increase the number of connections in PDF, DSR should be compared to AODV.

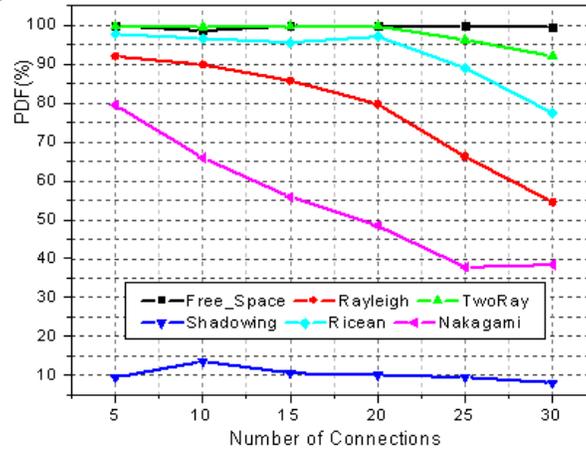

*Figure 2. a: AODV - PDF versus speed number of connections*

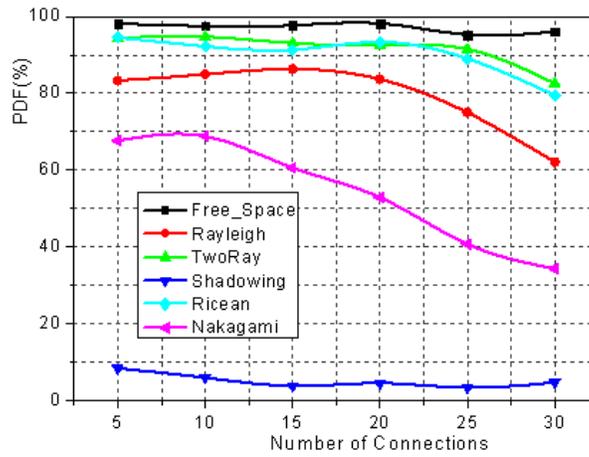

*Figure 2. b: DSDV - PDF versus speed number of connections*

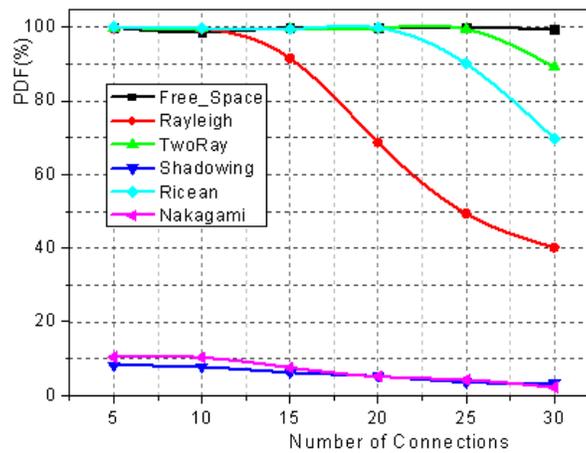

*Figure 2. c: DSR - PDF versus speed number of connections*

Meanwhile, we notice that the free-space and the two-ray ground deliver more packets than the other models, followed by first Rice model, second Rayleigh third Nakagami and finally the shadowing. Rice's model performance operates according to straight sight and employs the free-space for long distance prediction. Whereas, the shadowing bad performance is due to the low intensity of the signal caused by the obstacles. This results in the packet loss on weak links, displays wrongly the links disconnection and leads to the interruption and thus the dire need to set up a new itinerary.

## 5.2. Average End-to-End Delay

In figure 3, as expected, the delay is higher for non direct-sight propagation models (NLOS). Moreover, as there are more deliveries, the average delay also increases. Consequently, the packets have to wait more in a stand by position. In term of delays, we can observe that DSDV and AODV are more efficient than DSR. We also notice that delays for the two protocols increase rapidly according to the number of connections because of the high traffic congestion in some areas of the ad hoc networks. This congestion is triggered by the following factors:

- The ad hoc network with a dynamic topology may become a traffic congestion
- Both DSDV and AODV are considered as the number of hops in the measurement of a route. Besides, each of them has no device to choose the routes so that the data traffic can be distributed equitably.

AODV's delay increases more slowly than that of DSDV. This is accounted for by its use of priority criteria where in the protocol packet is given priority. Hence, a protocol packet is always treated prior to any data packet even if it arrives later. On the other hand DSDV does not distinguish between the protocol packets and the data ones during the waiting phase. Instead all packets are treated according to their arrival ranking.

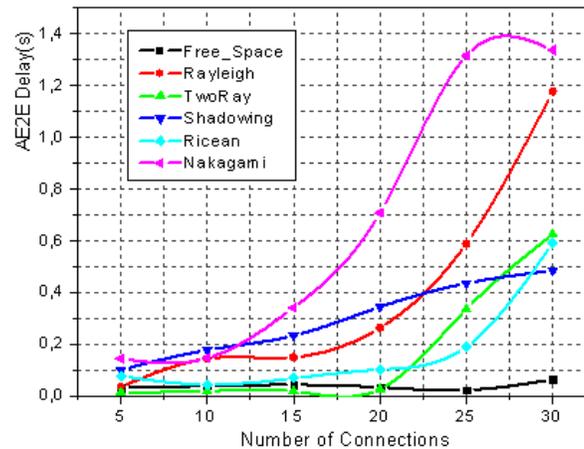

*Figure 3. a:  AODV- AE2E Delay versus number of connections*

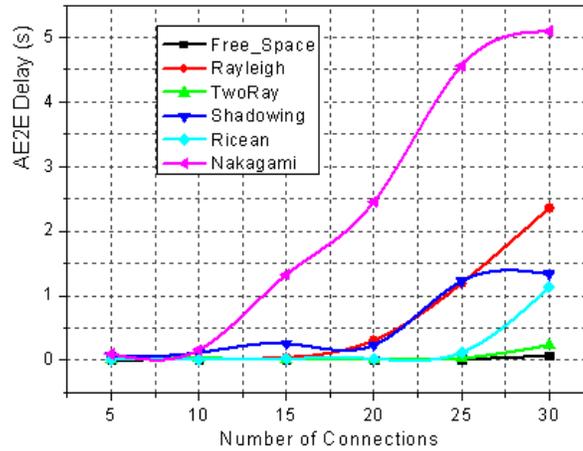

*Figure 3. b: DSDV- AE2E Delay versus number of connections*

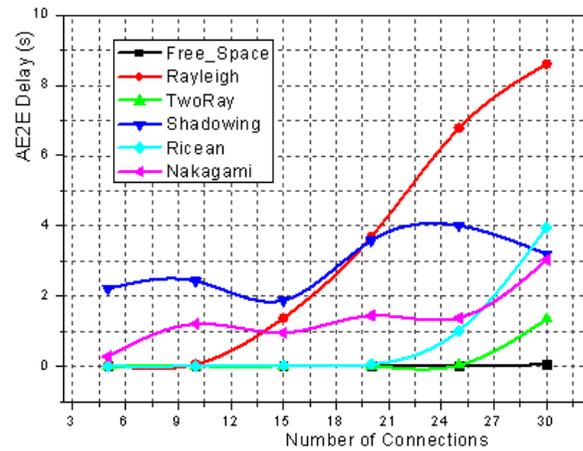

*Figure 3. c: DSR- AE2E Delay versus number of connections*

Similarly to PDF, we notice that the free-space and the two-ray ground endure less delay than the other models, followed by first Rice, second Rayleigh third Nakagami and finally the shadowing model. The weak performance of shadowing and Nakagami stems from the fact that when we observe the slope indicating the not mentioned collisions' rate, we realize that the phenomenon is accounted for.

## 5.3. Throughput:

In figure 4, we notice that the throughput diminishes significantly with an increase of the traffic load. DSDV protocol is steadier than AODV for the increasing number of connections. We also observe that the free-space model and the two-ray ground provide better throughput values than the other models.

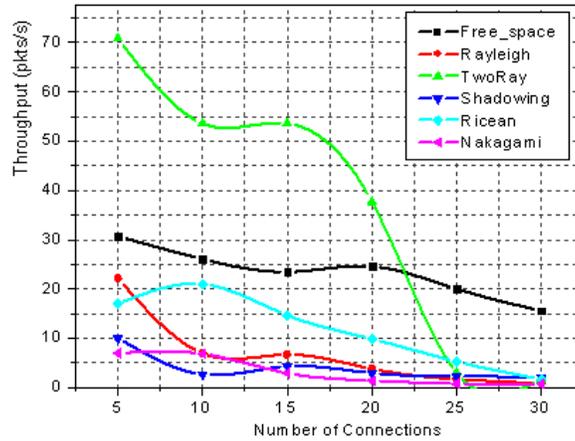

*Figure 4. a:  AODV- Throughput versus number of connections*

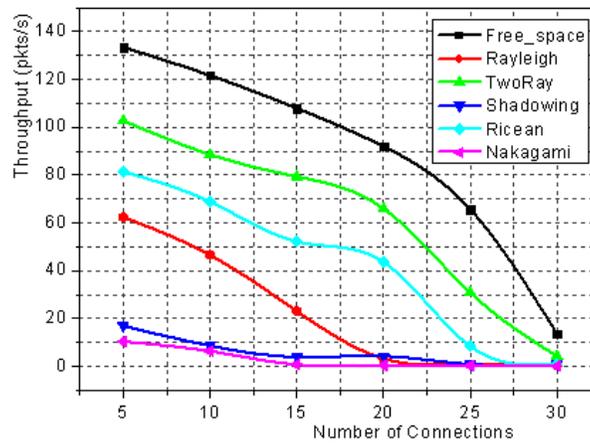

*Figure 4. b:  DSDV- Throughput versus  number of connection*

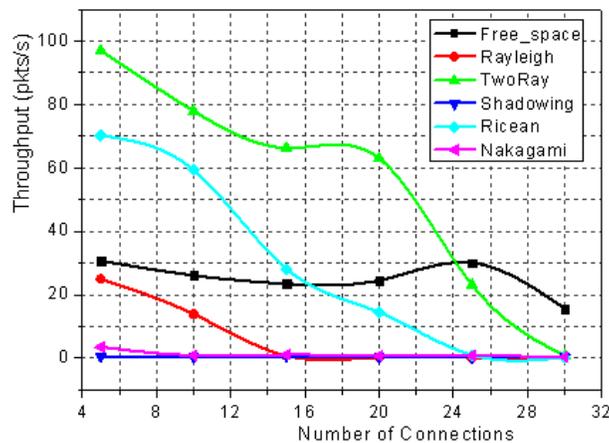

*Figure 4.c:  DSR- Throughput versus number of connection*

## 5.4. Minimal rates residual energy (MRRE)

We refer to minimal rates of nodes' residual energies in terms of their initial energies at the end of the simulation. The MRRE gives an idea about the protocol's tendency to maximize the nodes' life span and that of the whole network. Figure 5, shows the evolution of the whole consumed energy by varying the number of connections in 5, 10, 15, 20, 25 and 30. The energy consumption, the routing protocols AODV, DSR and DSDV also increase according to the number of sources.

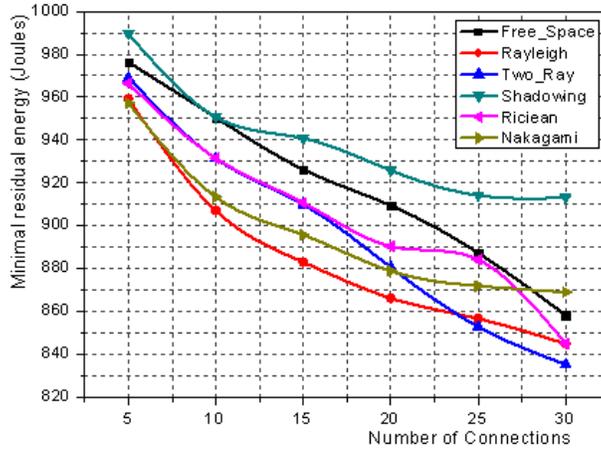

*Figure5. a: AODV- Minimal residual energy versus number of connections*

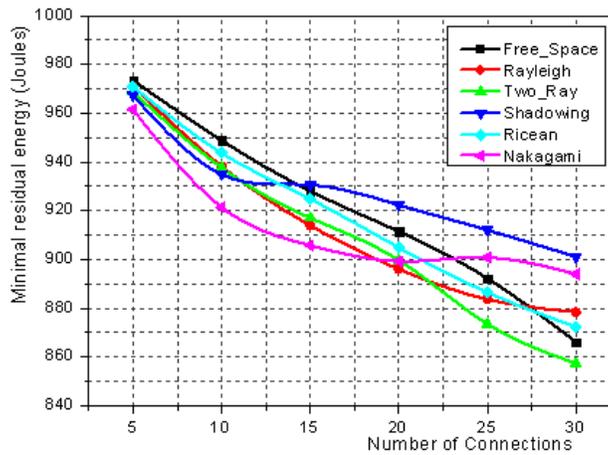

*Figure5. b: DSDV- Minimal residual energy versus number of connections*

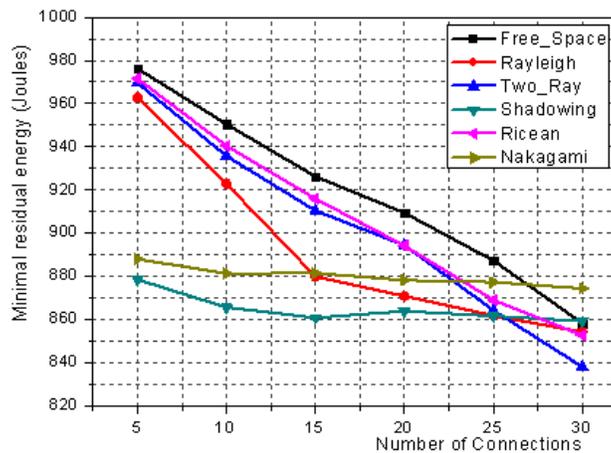

*Figure5. c: DSR- Minimal residual energy versus number of connections*

DSDV uses up less energy than AODV and DSR for each of radio propagation models. Thus, the DSDV routing protocol endows the networks with a longer life span than the two other protocols because the packet energy consumption is less then the latter. Since AODV and DSR create routes when asked to, they use up more energy then DSDV where the constructed table is updated once there is any change in neighboring topology. AODV, DSR and DSDV's life spans decrease when the number of connection increases because the collision rate and the retransmission tentative increase for an important traffic rate. Our study is based on the assessment of six propagation models such as; the free space, the two-Ray ground, shadowing, Rice's and Nakagami's models, and their impact on AODV, DSR and DSDV. The analysis of simulation results indicates that for different propagation models, the energy consumption is proportional to number of sent packets, packet delivery ratio and packet routing overhead. Thus the choice of the propagation model plays a key in the selection of the routing protocol because it has a serious impact on its performances.

## 6. CONCLUSIONS & PERSPECTIVES:

In this article, we study the impact of different radio propagation models on the performance of ad hoc networks. According to the simulation findings, we may state that the choice of the propagation models has a great impact on the routing protocol's performance. In this respect, we have identified both the determinist and the statistic modelizations. The simulation findings have revealed that the different propagation models have a considerable impact on the performance of the ad hoc mobile network. The latter decreases rapidly when the fading models, mainly Ricean, Rayleigh, Shadowing and Nakagami have been taken into consideration. The main reasons of their deterioration are the outcome of the big variation in the received intensity signal. According to the results to the routing protocols' performance, we find out that there is no preferable protocol among the others all scenarios and the assessing criteria. On the other hard, no matter how many connections there are, we notice that DSDV and AODV have a better delay term than DSR. Moreover, AODV and DSR have better performances in terms of delivery packet fractions. DSR uses the hidden memory to detect routes. In fact, this mechanism reduces these performances in terms of delay because of the abusive use of the hidden memory and the inability to delete the add routes. Nonetheless, it seems that the memory allows DSR to keep a weak overload. As far as the energy consumption is concerned, it is related to the number of treated packets and to the type of treatment. DSDV consumes less energy than AODV and DSR. Hence, we can state that DSDV routing protocol provides the network with a longer life span than the two other protocols. Since AODV and DSR create routes when asked to, they use up more energy then DSDV where the constructed table is updated once there is any change in neighboring topology.

To conclude, the simulation findings are to be taken as a strong reference on the three routing protocols' behaviour; however, it shouldn't be considered as an exact representation of its behaviour and real environment because of several simulation constraints such as: the dimension of movement field of mobile nodes, the traffic type and the simulation timing. In the forthcoming studies, we will look at the routing protocols' behaviours in the multi-channel environment and/or multi-networks in order to determine the key parameters that have an impact on the protocols' choice. Besides, we will try to develop new protocols or alter the existing ones.

## 7. REFERENCES

[1] Fourie, A. and D. Nitch, "SuperNEC: antenna and indoor propagation simulation program," IEEE Antennas and Propagation. Mag, Vol. 42, No. 3, 31–48, June 2000.

[2] A. Rhattoy, M. Lahmer, A. Zatni, « Simulation de La Couche Physique Dans Les Réseaux Mobiles », RNIOA'08, 05-07 Juin 2008, Errachidia, Maroc.

[3] Vaughan, R.; Andersen, J.B.; Channels, Propagation and Antennas for Mobile Communications, IEE, 2003.

[4] R. L. Freeman, Fundamentals of Telecommunications, 2nd ed. Hoboken, New Jersey, USA: John Wiley & Sons, Inc., 2005.

[5] Theodore S. Rapport. Wireless Communications: Principles and Practice. Prentice Hall PTR, Upper Saddle River, NJ, USA, 2002.

[6] D. R. Smith, Digital Transmission Systems. Norwell, Massachusetts, USA: Kluwer Academic Publishers, 1993.

[7] T. S. Rapport, Wireless Communications, Principles and Practice, 2nd ed. Prentice Hall, 2001.

[8] http://www.cubinlab.ee.unimelb.edu.au/~jrid/Docs/Manuel-NS2/node196.html.

[9] Simon Haykin and Michael Moher. Modern Wireless Communications. Pearson Prentice Hall, 2005.

[10] B. Sklar, "Rayleigh Fading Channels in Mobile Digital Communication Systems Part I: M. Carvalho and J. J. Garcia-Luna-Aceves, "Modelling Single-Hop Wireless Networks under Rician Fading Channels," in 5th IEEE Wireless Communications and Networking Conference (WCNC '04), vol. 4, March 21-25 2004, pp. 219–224.

[11] Bernard Sklar. Rayleigh fading channels in mobile digital communication systems. IEEE Communications Magazine, 35(7):90–100, July 1997.

[12] http://dsn.tm.uni-karlsruhe.de/medien/downloads_old/Documentation-NS-2-80211Ext-2008-02-22.pdf

[13] Li Tang, Zhu Hongbo, "Analysis and Simulation of Nakagami Fading Channel with MATLAB", Asia-Pasific, CEEM' 2003, Nov, 4-7, 2003, pp. 490-494 Hangzhou, China

[14] Feeney, L.M., "A Taxonomy for Routing Protocols in Mobile Ad Hoc Networks", in SICS Technical Report T99:07, 1999

[15] S. Kumar Gupta & R. K. Saket, "Performance metric comparison of AODV and DSDV Routing protocols in manet using ns-2", IJRRAS, Vol7,Issue3, pp. 339-350 June 2011

[16] V. Ramesh, Dr. P. Subbaiah, N. Koteswar Rao and M. Janardhana Raju, "Performance comparison and analysis of DSDV and AODV for MANET," International Journal on Computer Science and Engineering, vol. 02, pp. 183-188, 2010.

[17] C. Bettstetter, H. Hartenstein and X. Perez-Costa, "Stochastic Properties of the Random-Way Point Mobility Model," *Wireless Networks*, pp. 555 – 567, Vol. 10, No. 5, September 2004.

[18] Juan Carlos Cano and Pietro Manzoni., "A Performance Comparison of Energy Consumption for Mobile Ad Hoc Network Routing Protocols", Proceeding of 8th International Symposium on Modeling, Analysis and Simulation of Computer & Telecommunication System 2000.

## Biographies:

Dr. A. Rhattoy was educated at the Paul Sabatier University France; He obtained a Ph. D at National Polytechnic Institute of Institute of Toulouse France (in National Engineering School of Electrotechnics, electronics, Computer Science, Hydraulics and Telecommunications) in 1997. He has been a teaching for 13 years. He is currently a Professor and the Head of computer department in Moulay Ismail University at Higher School of technology, Meknes, Morocco; He conducts his research and Teaches Information Systems and Telecommunications.

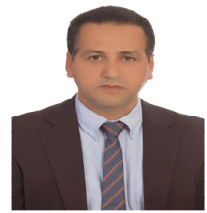

Dr. A. Zatni was educated at the Telecom Bretagne University France; He obtained a Ph. D at the National School of Engineers of Brest France in 1994. He has been teaching experience for 15 years. He is currently a Professor and the Head of computer science department in Ibnou Zohr University at Higher School of technology Agadir, Morocco, He conducts his research and teaches in computer science and Telecommunications .

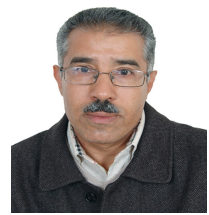